\documentclass[10pt, nobalancelastpage, twocolumn, floatfix, superscriptaddress, aps, prl]{revtex4-2}
\usepackage{amsmath, amssymb, graphicx, hyperref, xcolor, microtype}
\usepackage[export]{adjustbox}
\usepackage{braket}
\usepackage{enumitem}

\begin{document}

\title{Entanglement entropy and magic of ZX-diagrams}

\author{Marcin Szyniszewski}
\affiliation{Department of Computer Science, University of Oxford, Parks Road, Oxford OX1 3QD, UK}

\author{Razin A.\@ Shaikh}
\affiliation{Department of Computer Science, University of Oxford, Parks Road, Oxford OX1 3QD, UK}

\author{Aleks Kissinger}
\affiliation{Department of Computer Science, University of Oxford, Parks Road, Oxford OX1 3QD, UK}

\begin{abstract}
    Entanglement and non-stabilizerness are complementary resources underlying the complexity of quantum states, yet extracting either quantity from large quantum circuits generally requires exponentially scaling resources. In diagrammatic approaches to quantum computation, such as the ZX-calculus, both quantities likewise typically require expensive tensor contractions that are not native to graphical rewriting. Here we show that both entanglement and magic can be efficiently estimated directly from ZX-diagrams. Flow enables efficient extraction of a normal form separating a potentially extensively entangled graph-state backbone from non-Clifford Pauli gadgets. This structure yields additive upper and lower bounds on bipartite entanglement entropy, which we further tighten through a preprocessing procedure that removes or merges redundant non-Clifford contributions. The same normal form gives an upper bound on logarithmic stabilizer extent. We benchmark these methods in random unitary and monitored circuits, and a circuit combining Trotterized Hamiltonian evolution with Clifford layers, finding that the resulting bounds remain informative even at large qubit numbers, circuit depths, and internal spider counts. Our results provide scalable probes of entanglement and non-stabilizerness of ZX-diagrams and establish a bridge between ZX-calculus and quantum many-body resource characterization, with applications across quantum computing, quantum information, and condensed-matter physics.
\end{abstract}

\maketitle

\emph{Introduction.}---%
Entanglement and non-stabilizerness are central resources underlying the complexity of quantum states and dynamics. Entanglement entropy characterizes bipartite quantum correlations, distinguishing area-law, critical, and volume-law regimes in quantum many-body systems~\cite{Amico2008, Eisert2010, Hastings2007, Calabrese2004, Page1993, Nahum2017, Fisher2023}. Yet entanglement alone does not determine computational complexity: stabilizer states remain efficiently classically simulable even with extensive entanglement~\cite{Gottesman1998, Aaronson2004, Hein2004, Hein2006, Fattal2004}. The complementary resource of \emph{magic}, or non-stabilizerness, quantifies departure from stabilizer states and controls key costs of fault-tolerant computation and stabilizer-based classical simulation~\cite{Bravyi2005, Veitch2014, Howard2017, Bravyi2019, Seddon2019}. Magic has also emerged as a many-body observable with applications ranging from quantum criticality and tensor-network states to chaotic dynamics and monitored quantum circuits~\cite{White2021, Leone2022, Haug2023, Lami2023, Niroula2024, Bejan2024, Turkeshi2025}. Developing scalable methods that can access both resources is therefore important not only for quantum computation but also for characterizing strongly correlated quantum dynamics.

\begin{figure}[b]
    \centering
    \includegraphics[width=\columnwidth]{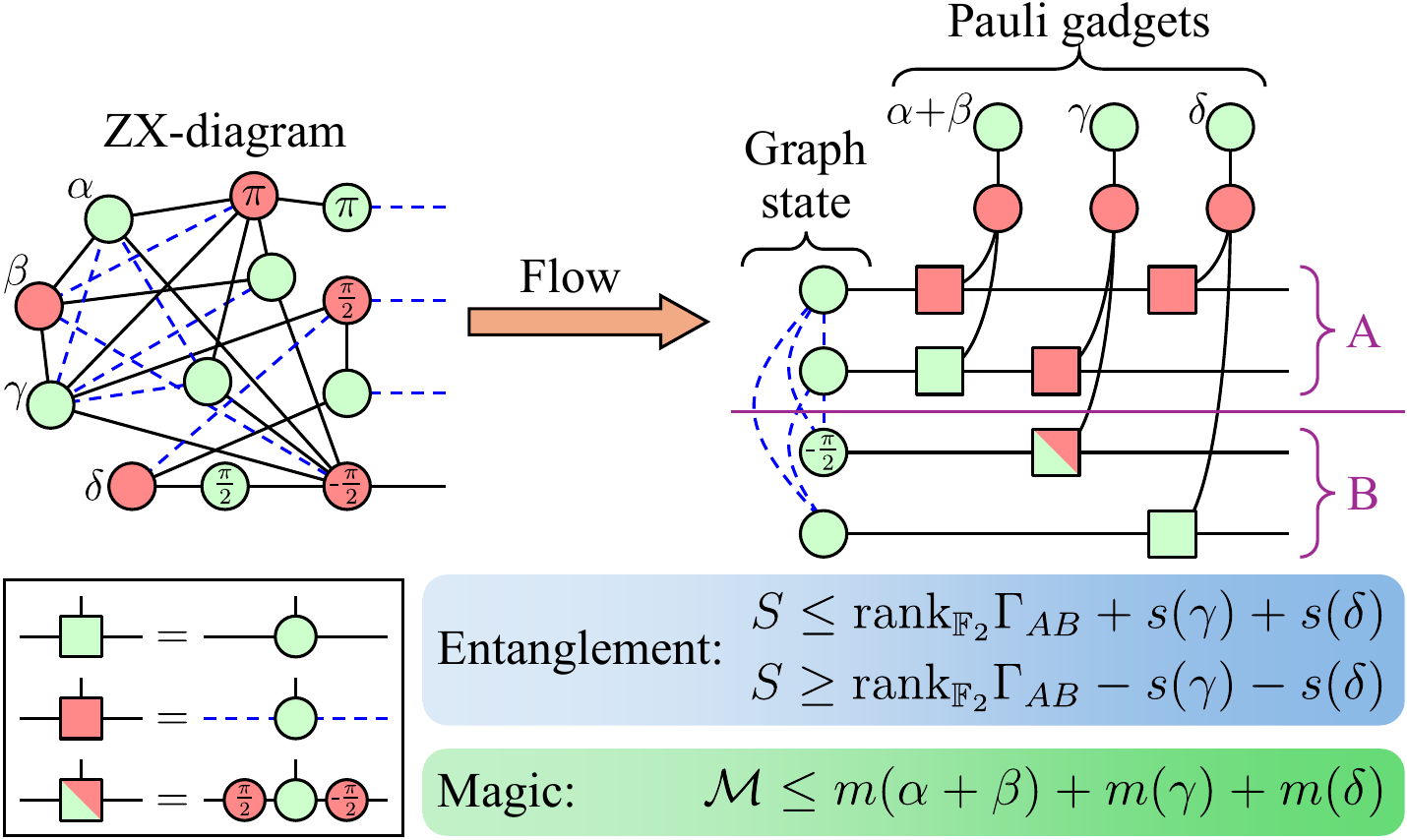}
    \caption{Estimating entanglement entropy and magic of ZX-diagrams. Flow enables conversion to a normal form separating graph-state backbone from Pauli gadgets. Gadget crossing $A|B$ bipartition add $s(\varphi) = h_2(\sin^2 \frac{\varphi}{2})$ to entanglement. All gadgets add $m(\varphi) = 2 \log_2 [ \sqrt{1{-}|\sin \varphi|} + \sqrt{1{-}|\cos \varphi|} ]$ to magic. ZX simplification can tighten both bounds.}
    \label{fig:summary}
\end{figure}

Extracting these quantities first involves selecting a relevant quantum state representation. Tensor-network methods exploit low-entanglement structure extremely effectively, but their computational cost is controlled by bond dimension or, more generally, by the width of the associated contraction problem~\cite{Verstraete2008, Schollwock2011, Orus2014, Markov2008}. Consequently, volume-law states and high-width networks can become prohibitively expensive to contract even for coarse resource estimates. The ZX-calculus may provide a natural setting for tackling this problem. Recent work has shown that ZX-diagrams (which can be viewed as a specific flavor of tensor networks) can be contracted efficiently when their rank-width is small~\cite{Kuyanov2026}. Here we pursue a complementary question: whether useful information about entanglement and magic can be extracted from a ZX-diagram without contracting the represented state, and without requiring its entanglement to be small.

The ZX-calculus is a complete graphical language for qubit quantum mechanics, in which linear maps (states, gates, measurements) are represented by ZX-diagrams and transformed by exact rewrite rules~\cite{Coecke2011, Backens2014, Backens2014CT, Jeandel2018, Wetering2020}. Beyond its established applications to circuit optimization, verification, and classical simulation~\cite{Duncan2020, Kissinger2019, Kissinger2020PyZX, Backens2021, Kissinger2022Classical, Sutcliffe2024}, ZX-based methods have recently begun to find applications in quantum many-body physics, including AKLT and topological states~\cite{East2022, MasMendoza2025, Lu2025}, measurement-induced transitions~\cite{BuznachAhituv2025}, gauge theories~\cite{Gorantla2025, Wong2026}, quantum many-body scars~\cite{Szyniszewski2026}, entanglement membranes~\cite{Sommers2024, Ha2025}, and exact magic dynamics in interacting circuits~\cite{Lopez2024, Dowling2025}. These developments suggest that graphical rewriting can expose physical structure obscured in conventional circuit or tensor-network representations. In particular, graph-like ZX-diagrams~\cite{Duncan2020} expose an underlying stabilizer structure, while flow conditions enable efficient conversion to circuit and Pauli-exponential representations~\cite{Browne2007, Simmons2021, Backens2021, Kissinger2026}.

In this Letter, we develop scalable methods for estimating entanglement and magic directly from ZX-diagrams without tensor contraction (see Fig.~\ref{fig:summary}). We exploit flow to efficiently extract a normal form that separates the stabilizer part from the non-Clifford unitaries. This structure allows upper- and lower-bounding entanglement entropy. We introduce an algorithm that simplifies and reduces the non-Clifford gadgets that contribute to the entropy calculation, allowing for tighter theoretical estimates. We also provide an upper bound on logarithmic stabilizer extent~\cite{Bravyi2019, Seddon2019, Heimendahl2021, Seddon2021}. The resulting bounds on entanglement and magic are additive over the non-Clifford contributions, become exact in the Clifford limit, and according to our numerical tests, can remain informative even when the underlying graph state has extensive entanglement, and the number of qubits, the circuit depth, or the number of internal ZX spiders are large.

\emph{Flow criteria expose the normal form.}---%
We first introduce the notation for ZX-diagrams, which represent qubit linear maps in ZX-calculus. The diagrams are generated by Z-spiders, \raisebox{-0.3\totalheight}{\includegraphics{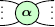}} $= \ket{0 \cdots 0} \bra{0\cdots0} + e^{i\alpha} \ket{1\cdots1} \bra{1 \cdots 1}$, and X-spiders, \raisebox{-0.3\totalheight}{\includegraphics{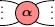}}, defined analogously in the X basis. Spider phases $\alpha \in [0, 2\pi)$ are called Clifford if they are integer multiples of $\pi/2$. It is convenient to define a Hadamard edge, 
\raisebox{-0.3\totalheight}{\includegraphics{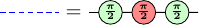}}.
%$\begin{ZX}[column sep=1cm] \ar[r,blue,dashed] & \end{ZX} = \begin{ZX} \ar[r] & \zxFracZ{\pi}{2} \rar & \zxFracX{\pi}{2} \rar & \zxFracZ{\pi}{2} \rar & \end{ZX}$.
More complicated diagrams are constructed by connecting these generators, corresponding to composition and tensor product of their associated linear maps.

Consider an $L$-qubit state-preparation ZX-diagram and seek the normal form
\begin{equation}
    \ket{\psi} = U_N \cdots U_1 \ket{G}, \qquad U_i=e^{-i \varphi_i P_i / 2},
    \label{eq:normalform}
\end{equation}
where $\ket{G}$ is a graph state and $U_i$ are Pauli gadgets~\cite{Cowtan2020} (exponentials of Hermitian Pauli strings $P_i$ with phases $\varphi_i$, also called Pauli exponentials, or Pauli product rotations). This form is well adapted to entanglement and magic estimation, as it separates the stabilizer backbone, whose entanglement contribution is graph-theoretic, from the non-Clifford deformations, encoded by arbitrary-angle Pauli exponentials. Importantly, these deformations are unitary: those confined to one side of the bipartition leave the entanglement unchanged, while those crossing it contribute only a bounded amount, as discussed later. Nonunitary deformations (e.g., weak postselected measurements) can instead change entanglement extensively even when confined to one subsystem~\cite{Bennett1996, Vidal1999, Lo2001}.

From the ZX-calculus perspective (see Fig.~\ref{fig:summary}), $\ket{G}$ is easily represented by a ZX-diagram: up to local Clifford operations, Clifford-phase Z-spiders representing the qubits are connected by Hadamard edges according to the underlying graph. 
Stabilizer generators of $\ket{G}$ can then be read directly as $\mathcal{S}_v = X_v \prod_{v'\in N(v)} Z_{v'}$, where $N(v)$ denotes all Z-spiders connected to spider $v$ by Hadamard edges. Its bipartite entanglement entropy $S_0 = \text{rank}_{\mathbb{F}_2} \Gamma_{AB}$ remains efficiently computable even when extensive; $\Gamma_{AB}$ is the adjacency submatrix across the cut $A|B$, with rank calculated over the two-element field $\mathbb F_2$~\cite{Hein2004, Hein2006, Fattal2004}. Pauli gadgets also have a natural ZX representation containing a Z-spider with a potentially non-Clifford phase $\varphi_i$. Pauli gadgets are widely used in ZX-based circuit optimization, where they capture multi-qubit parity-phase operations~\cite{Kissinger2019, Cowtan2020}; ZX-calculus also explicitly treats Pauli exponentials as universal circuit-level primitives~\cite{KissingerWetering2024Book}.

This normal form is expressive enough to describe any arbitrary pure state, yet finding such a decomposition for a generic ZX-diagram need not be efficient~\cite{debeaudrap2022extraction}. Flow criteria provide a structural condition that makes this conversion polynomial in the size of the ZX-diagram. For graph-like diagrams, Pauli flow gives a causal structure from which one can efficiently extract a circuit~\cite{Simmons2021, Backens2021}. ZX-flow extends this notion to arbitrary ZX-diagrams and is equivalent, up to Clifford rewrites, to Pauli flow on a graph-like diagram~\cite{Kissinger2026}. Crucially here, diagrams with flow can be turned into a Clifford isometry followed by a sequence of Pauli exponentials, providing a polynomial-time route to Eq.~\eqref{eq:normalform} for states and the subsequent entanglement and magic bounds.

\emph{Entanglement entropy bounds for ZX-diagrams.}---%
General bounds on entanglement generation based on operator Schmidt rank or entangling power~\cite{Linden2006, Gour2007} do not exploit the detailed composition of Pauli gadgets. Sharper additive bounds are known for individual Pauli gadgets applied to arbitrary states~\cite{Leifer2003, Dur2001, Linden2009, Kraus2001, Chefles2005, Berry2003, Wang2003} and depend explicitly on the gadget angles $\varphi_i$, providing the natural benchmark for our construction (see Appendix~A in Supplemental Material, SM~\cite{SuppMat}).

Here, assuming no coherent branch interference defined later, we obtain simpler and tighter bounds on von Neumann entanglement entropy $S$ for a bipartition $A|B$, defined as $S=-\text{tr}_A (\rho_A \log_2 \rho_A)$ for a reduced density matrix $\rho_A = \text{tr}_B (\ket{\psi} \bra{\psi})$. The bounds (see Appendix~B, SM~\cite{SuppMat}) for a state in normal form Eq.~\eqref{eq:normalform} are given by
\begin{subequations}
    \begin{align}
    S &\le 
        \text{rank}_{\mathbb F_2}\Gamma_{AB}
        +\sum_{i\in\mathcal{C}} h_2\!\left(\sin^2\!\frac{\varphi_i}{2}\right),\\
    S &\ge
        \text{rank}_{\mathbb F_2}\Gamma_{AB}
        -\sum_{i\in\mathcal{C}} h_2\!\left(\sin^2\!\frac{\varphi_i}{2}\right),
    \end{align}\label{eq:mainbound}%
\end{subequations}
where $\text{rank}_{\mathbb F_2} \Gamma_{AB}$ is the entanglement entropy of the graph state $\ket G$, $h_2(x)=-x\log_2 x-(1-x)\log_2(1-x)$ is the binary entropy, and $\mathcal{C}$ is a set of gadgets crossing the bipartition $A|B$. Additionally, entanglement is always trivially bounded as $0 \le S \le \min(|A|,|B|)$. Our bounds are additive w.r.t.\@ Pauli gadgets and remain tight even after many weak non-Clifford rotations, which should be contrasted with the generic entangling-capacity bounds~\cite{Leifer2003}: we find an improvement from linear to quadratic-logarithmic scaling for small phases $\varphi_i$, allowing informative estimates at depths where generic bounds can already saturate the trivial interval.

Analogous bounds hold for R\'enyi entropies $S_\alpha$ where $\alpha>1/2$ is the R\'enyi index: replace $h_2(x)$ by the binary R\'enyi entropy $h_2^{\alpha}(x) = (1-\alpha)^{-1} \log_2 (x^\alpha + (1-x)^\alpha)$ in the upper bound and by $h_2^{\alpha'}(x)$ with conjugate index $\alpha'=1/(2-1/\alpha)$ for the lower bound. The bounds are therefore generally asymmetric, recovering the symmetric form only for the von Neumann entropy, $\alpha=1$.

The proof outline is as follows. Each Pauli gadget decomposes into two branches (sine and cosine), while for every Pauli string $P$, $P\ket G$ has the same flat Schmidt spectrum as $\ket G$. The upper bound follows by dephasing the superposition into sectors and contributes at most the Shannon entropy of the branch distribution to the entanglement entropy. The lower bound follows from an entropic uncertainty relation~\cite{Maassen1988} between a complete stabilizer-orbit basis and the product basis containing the final state's Schmidt vectors.

Both bounds hold only when branches do not coherently interfere, $P_{i_1} \cdots P_{i_k} \neq \gamma \, \text{Stab} (|G\rangle)$ for any subsequence of Pauli gadgets, with $\gamma = \pm 1$ for even $k$, and $\gamma = \pm i$ for odd $k$. For example, two commuting gadgets with branch interference can be replaced by a single gadget with phase $\varphi_1 \pm \varphi_2$ (a ZX-calculus rewrite rule known as gadget fusion). To reduce coherent interferences and tighten our bounds, we introduce a preprocessing procedure. First, gadgets are fused whenever circuit commutation exposes identical Pauli supports; this is standard in ZX-based optimization and can be automated, for example, with PauliOpt~\cite{Gogioso2023, MeijervandeGriend2025, Winderl2023}. Second, within a commuting region, we also fuse two gadgets whose Paulis differ by a Pauli stabilizer of the state immediately before that region. Third, Pauli gadgets whose phases are outside $\varphi\in[-\pi/4,\pi/4]$ have their residual Clifford part unfused and pushed to the graph state. Fourth, if multiplying a Pauli gadget by a stabilizer of the immediately preceding state produces a Hermitian Pauli supported entirely in one subsystem, that gadget can be excluded from $\mathcal C$.

Our scalable preprocessing searches only for pairwise fusions in commuting regions and therefore does not certify the bounds in Eq.~\eqref{eq:mainbound} for arbitrarily large diagrams: sufficiently many gadgets necessarily cause coherent interference of branches. What is suppressed for generic large systems is the occurrence of low-weight interferences, and interferences involving $k$ small-angle gadgets enter only through $k$th-order interference amplitudes. Thus, we expect our preprocessing to perform well in unstructured circuits (e.g., random circuits) but be more vulnerable to repeated structure (e.g., Trotterized evolution).

\begin{figure}
    \centering
    \includegraphics[width=\columnwidth]{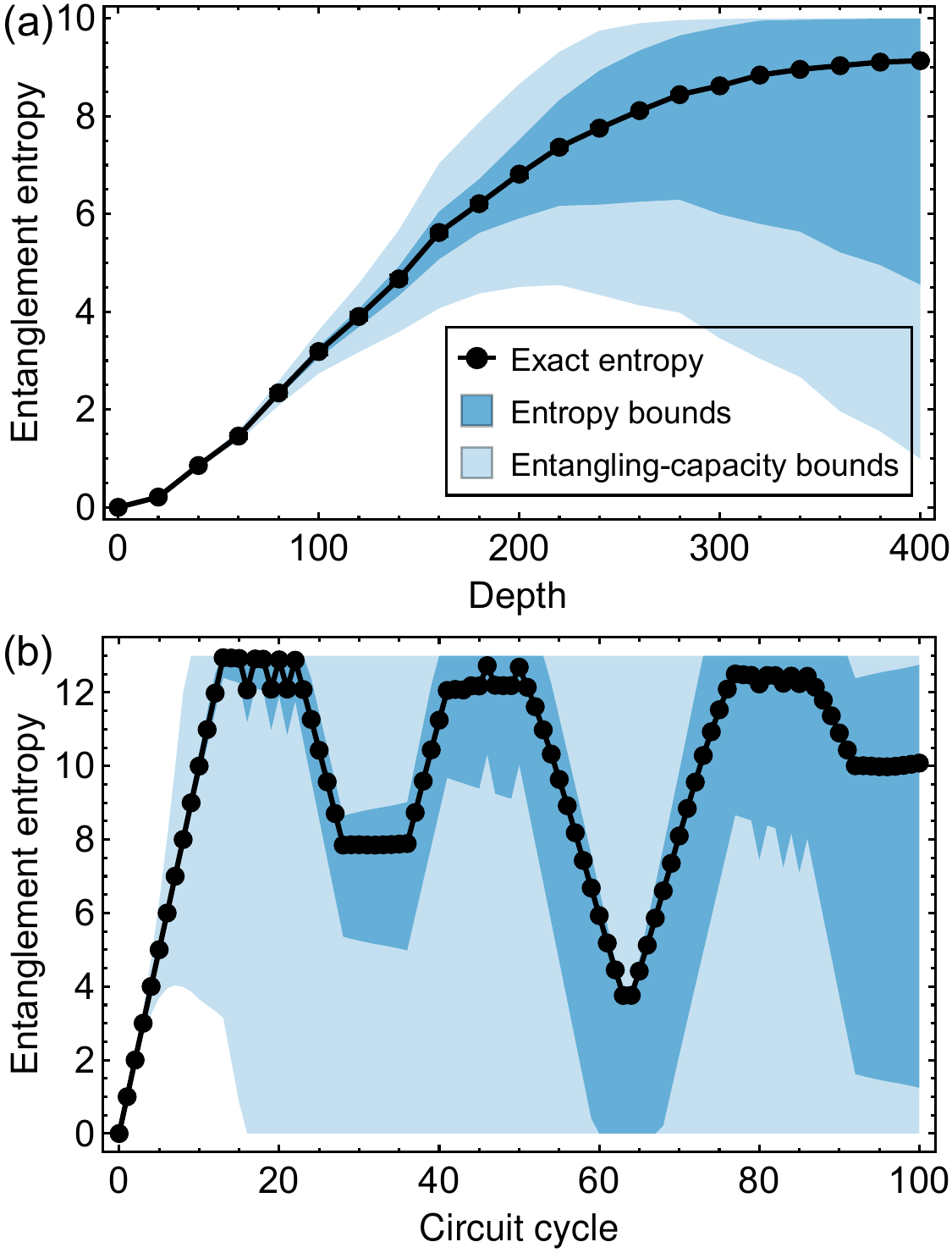}
    \caption{Half-chain entanglement entropy in (a)~random Clifford-plus-small-phase circuit for $L=20$ qubits, and (b)~circuit combining Trotterized Hamiltonian evolution with Clifford layers for $L=26$ qubits. The entropy bound from Eq.~\ref{eq:mainbound} remains close to the exact entropy (darker blue region), while generic entangling-capacity bounds broaden rapidly (lighter blue region).}
    \label{fig:entropy_bounds}
\end{figure}

\emph{Numerical tests.}---%
We perform example simulations using PyZX~\cite{Kissinger2020PyZX} to construct and simplify ZX-diagrams, and PauliOpt~\cite{Gogioso2023, MeijervandeGriend2025, Winderl2023} for Pauli gadget manipulation. All circuits are explicitly shown in Appendix D, SM~\cite{SuppMat}.

We first consider random circuits on $L=20$ qubits composed of CNOT, Hadamard, $S$ phase gates, and single-qubit $Z$-phase gates, with the non-Clifford phases chosen independently from a small-angle distribution. Specifically, the circuit is initialized in an all-zero state, and for each time step we randomly choose a CNOT gate with probability $0.3$, and a single-qubit gate from the set $\{H, S, Z_\alpha\}$ with probability $0.7$, where $\alpha\in[0,\pi/10]$. We evolve to depth $T=400$ (which here specifies the number of gates) and average over $300$ circuit realizations. Fig.~\ref{fig:entropy_bounds}(a) compares the exact half-chain entropy with Eq.~\eqref{eq:mainbound} and generic entangling-capacity bounds, both after preprocessing. Our estimates track entanglement growth over a substantially longer interval, while the generic bounds broaden rapidly. Among all samples, spanning different random realizations and circuit depths, we found no violation that falls outside Eq.~\eqref{eq:mainbound}, which would suggest branch interference.

To test our bounds under both substantial entanglement growth and repeated circuit structure, we consider a circuit combining Trotterized Hamiltonian evolution with Clifford layers on $L=26$ qubits. Starting from $\ket{+-+-\cdots}$ state, each cycle consists of: one layer of $S$ gates, one brickwork layer of CNOT gates, and Trotterized evolution operator $e^{i dt H}$, where $H = \sum_i(J_x X_i X_{i+1} + J_y Y_i Y_{i+1} + J_z Z_i Z_{i+1} + h_z Z_i + h_x X_i)$, with $dt = 0.01$ and $J_x = 1.0, J_y = 0.8, J_z = 0.5, h_z = 0.3, h_x = 0.2$ chosen generically. Repetition makes this setting unfavorable for any preprocessing restricted to low-order gadget fusions. Nevertheless, Fig.~\ref{fig:entropy_bounds}(b) shows that the generic bounds quickly become uninformative, filling in the trivial interval $[0, L/2]$, whereas Eq.~\eqref{eq:mainbound} continues to track the exact entropy closely for much larger times. The first observed violation of the bound occurs at step $64$, which is the periodicity of the Clifford circuit, likely signaling when the higher-order branch interferences become important.

\emph{Magic upper bound for ZX-diagrams.}---%
Magic controls the overhead required to represent or sample non-Clifford dynamics~\cite{Beverland2020, Seddon2019, Heinrich2019} and provides an experimentally accessible diagnostic of genuinely non-Clifford quantum resources in many-body platforms~\cite{Oliviero2022}.
We quantify magic using the logarithmic stabilizer extent~\cite{Bravyi2019, Heimendahl2021, Seddon2021}, which for a pure state $\ket{\psi}$ is defined as
\begin{equation}
    \mathcal{M}(\ket{\psi})=\log_2 \! \Big[ \min_{\ket{\psi}=\sum_\alpha\!c_\alpha \ket{\phi_\alpha}} \!\Big(\sum_\alpha |c_\alpha|\Big)^2\Big] ,
\end{equation}
where the minimization is over all decompositions of $\ket{\psi}$ into stabilizer states $\ket{\phi_\alpha}$. The log stabilizer extent is non-negative, vanishes on stabilizer states, and characterizes the cost of stabilizer-based classical simulation.

\begin{figure}
    \centering
    \includegraphics[width=\columnwidth]{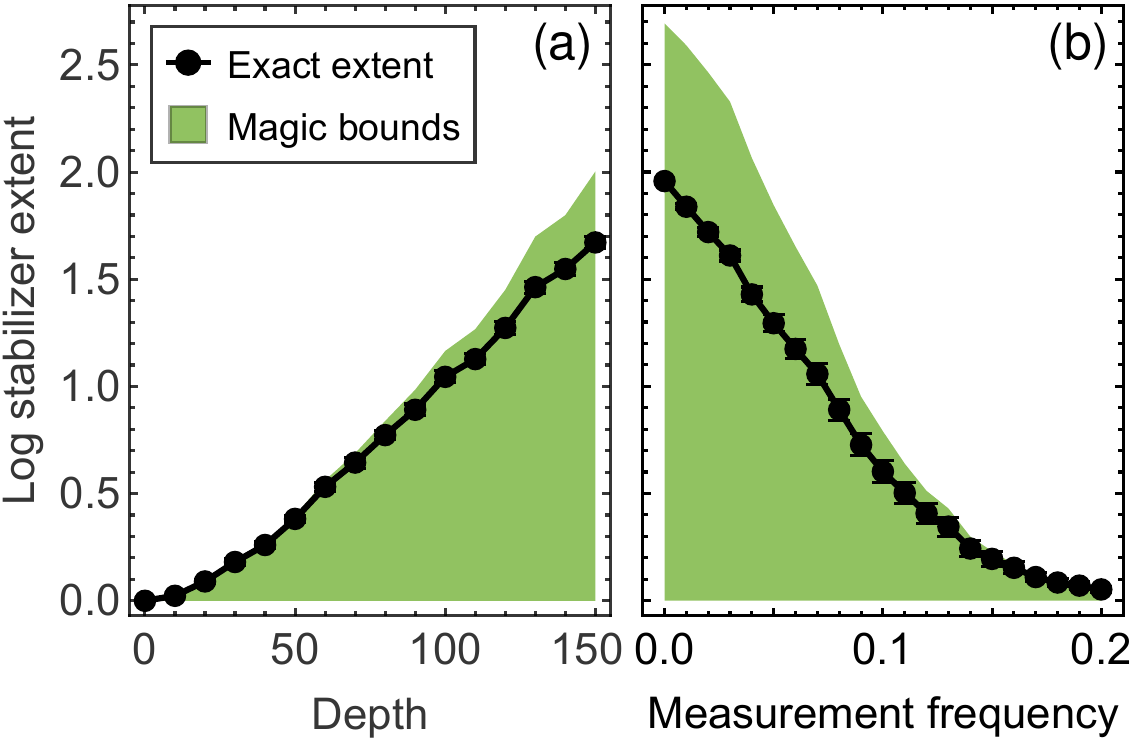}
    \caption{Logarithmic stabilizer extent for (a)~random Clifford-plus-small-phase unitary circuit and (b)~random circuit with mid-circuit measurements, with $L=8$ qubits. The magic bound Eq.~\ref{eq:magic_bound} is shown in green.}
    \label{fig:magic_bound}
\end{figure}

The decomposition in Eq.~\eqref{eq:normalform} naturally separates the stabilizer resource from non-stabilizerness and allows upper-bounding $\mathcal{M}$ (see Appendix~C, SM~\cite{SuppMat}),
\begin{equation}
    \mathcal{M}(\ket{\psi}) \le 2 \sum_{i\in \mathcal{O}}
    \log_2\! \left[ \sqrt{1{-}|\sin \varphi_i|} + \sqrt{1{-}|\cos \varphi_i|} \right]\!.
    \label{eq:magic_bound}
\end{equation}
Unlike the entropy bounds, this result requires no assumptions on branch interference, yet it has similar practical advantages: it is additive over gadgets and efficiently computable without tensor contraction or optimization over stabilizer decomposition. Moreover, it vanishes at Clifford angles and becomes tight near the stabilizer limit. Thus, small non-Clifford deformations generate only a controlled amount of magic, with the above expression giving an efficient, diagrammatic upper bound on the corresponding non-stabilizerness. The sum includes all gadgets $\mathcal{O}$ that meaningfully impact the state (Pauli gadgets with supports being stabilizers of the state are excluded).

We benchmark Eq.~\eqref{eq:magic_bound} against exact $\mathcal M$ computed using Ref.~\cite{Hamaguchi2025} and its accompanying code. The test is performed on a random circuit introduced in the entanglement numerics, and a random circuit with mid-circuit measurements, both for $L=8$ and 100 random realizations. The monitored circuit is initialized in the all-plus state and evolves for $T=L$ steps, each comprising: a brickwork layer of CNOT gates, a layer of independent Z-phase gates with phases uniform on $[0, \pi / 10)$, and a layer of Z measurements (postselected onto $\ket{0}$) applied stochastically with frequency $p$ per qubit. Small phases allow probing the accumulation of weak non-Clifford rotations without rapidly saturating magic in this small system. Despite the measurements, the ZX-diagram of this circuit possesses flow and thus can be transformed to the normal form Eq.~\eqref{eq:normalform}: strings of Pauli-Z propagate through CNOTs and commute with the Z-phase gates and Z measurement projectors, allowing the non-Clifford rotations to be moved near the outputs. Additionally, flow-preserving ZX simplification~\cite{Kissinger2019} can tighten the estimate before extraction by disconnecting diagram components (measurements can be thought of as breaks in the diagram), making this a particularly compelling test.

In both random ensembles shown in Fig.~\ref{fig:magic_bound}, the exact magic nearly saturates the bound, likely due to the random nature of the circuits. Increasing $p$ in the monitored circuit produces a crossover from a highly magical state to a near-stabilizer state: in fact, the computational all-zeros state is absorbing in this dynamics. These results motivate using the scalable estimate to investigate magic survival and purification, including possible signatures of measurement-induced magic transitions in related monitored circuits~\cite{Bejan2024}.

\emph{Discussion.---}%
We have shown that entanglement and non-stabilizerness can be estimated directly from the structure of ZX-diagrams without contracting the represented quantum state. The key step is provided by flow conditions, giving an efficient decomposition into a graph-state backbone dressed by Pauli gadgets, which separates two distinct sources of quantum complexity: the graph state may carry extensive entanglement while remaining entirely stabilizer, whereas non-Clifford gadgets encode deviations from stabilizer structure. We further introduce a preprocessing procedures that substantially tighten the entanglement bounds and the upper bound on magic. Our additive bounds are efficiently calculable even for large diagrams and remain informative for large qubit numbers and circuit depths.

Our research opens up many future research directions across different related disciplines. The methods provided here should find use in ZX-based compilation, as efficient estimates for entanglement and magic could be used in algorithms using resource-aware cost functions. On the algorithmic side, further improvements could also be made to preprocessing using methods to remove higher-order gadget fusions. Perhaps such fusions can be formulated as low-weight dependencies in the quotient of the Pauli group by the state stabilizer, suggesting coding-theoretic or matroid methods that could be substantially cheaper than exhaustive subset search. Repeated Trotter and Floquet structures should be particularly amenable to such targeted searches.

This construction could potentially be generalized beyond pure state-preparation diagrams with flow, for example to mixed states or trajectories of monitored dynamics. The latter would be particularly interesting in the context of measurement-induced entanglement and magic transitions~\cite{BuznachAhituv2025, Bejan2024}, where the two resources can display distinct dynamical behavior. Extending our bounds to subregions could provide a diagrammatic probe of entanglement and magic spreading fronts. Related work has already shown that ZX-calculus can yield exact stabilizer R\'enyi entropies in special many-body circuits~\cite{Lopez2024} and reveal hidden structure in monitored Clifford transitions~\cite{BuznachAhituv2025}.

More broadly, ZX-calculus has largely developed as a language for manipulating quantum computations, whereas entanglement entropy is one of the central quantities in quantum many-body theory and condensed matter physics. By bridging the two, our research opens the possibility of using diagrammatic rewriting of ZX-calculus for exposing and controlling the entanglement structure of large physically-relevant quantum many-body systems.

\emph{Acknowledgments.---}%
A.~K. and M.~S.\@ were supported by the Engineering and Physical Sciences Research Council grant number EP/Z002230/1: (De)constructing quantum software (DeQS). R.~S.\@ was supported by the Clarendon Fund Scholarship. The authors acknowledge the use of the University of Oxford Advanced Research Computing (ARC) facility in carrying out this work~\cite{Richards2015}.

\bibliography{refs}

\widetext
\clearpage

\setcounter{secnumdepth}{1}
\renewcommand{\theequation}{S\arabic{equation}}
\renewcommand{\thefigure}{S\arabic{figure}}
\renewcommand{\thesection}{S\arabic{section}}
\setcounter{equation}{0}
\setcounter{figure}{0}

\begin{center}
    \large \textbf{Supplemental Material to ``Entanglement entropy and magic of ZX-diagrams''}
\end{center}

\appendix

\section{Maximum entanglement entropy change for a Pauli gadget on an arbitrary state}

Consider applying a Pauli gadget $e^{-i \phi P/2}$ to an arbitrary pure state $\ket{\psi}$ with entanglement entropy $S_0$ across the bipartition $A|B$, and asking what the maximal change in the entropy is~\cite{Leifer2003, Dur2001, Linden2009, Kraus2001, Chefles2005, Berry2003}. We can first simplify the problem by turning it into a 2-qubit problem. One can transform Pauli strings $P_A$ and $P_B$ in each subsystem into a single $Z$ Pauli operator using Clifford unitaries that act only within each subsystem. Hence, the problem is equivalent to determining the maximum change in entanglement entropy of a 2-qubit state (possibly with ancillas) after applying a phase gadget $e^{-i \phi Z \otimes Z/2}$.

For a 2-qubit pure state $\ket{\psi} = a\ket{00} + b\ket{01} + c\ket{10} + d\ket{11}$, the entanglement entropy is directly related to concurrence $C=2|ad-bc|$ as
\begin{equation}
    S = h_2 \!\left(\frac{1+\sqrt{1-C^2}}{2}\right),
\end{equation}
where $h_2(x)=-x \log x-(1-x)\log(1-x)$ is the binary entropy. Concurrence can be parametrized by angle $C=|\sin(\theta/2)|$, and thus the initial entanglement entropy has a form
\begin{equation}
    S(\theta) = h_2 \!\left(\frac{1+|\cos(\theta/2)|}{2}\right).
\end{equation}

After one applies the phase gadget $e^{-i \phi Z \otimes Z/2}$, the coefficients of the state change as $a\mapsto e^{-i \phi/2}a, b\mapsto e^{i \phi/2}b, c\mapsto e^{i \phi/2}c, d\mapsto e^{-i \phi/2}d$. The concurrence is now
\begin{equation}
    C^2 = 4(|a d|^2+|b c|^2-2|a d|\,|b c|\cos \delta), 
    \quad
    \delta = \arg(ad) - \arg(bc) - 2 \phi.
\end{equation}
Differentiating gives
\begin{equation}
    \Big| \frac{d}{d\phi} C^2 \Big| = 16 |ad|\, |bc|\, |\sin \delta| \le 2 C\sqrt{1-C^2}.
\end{equation}
The inequality can be checked easily using normalization and $2|ad| \le |a|^2+|d|^2, 2|bc| \le |b|^2+|c|^2$. Since $C=|\sin(\theta/2)|$,
\begin{equation}
    \frac{d\theta}{d\phi} = \frac{d(C^2)/d\phi}{C\sqrt{1-C^2}} \le 2.
\end{equation}
Integrating from 0 to $\phi$, gives $|\theta_\phi-\theta_0|\le 2|\phi|$, and hence the phase gadget can move the angle $\theta$ by at most $2|\phi|$.

Hence, the largest possible entropy change is bounded by the largest change over an angular displacement of $2|\phi|$~\cite{Leifer2003},
\begin{equation}
    |\Delta S| \le 
    \max_\theta \Big[ S(\theta+2\phi) - S(\theta) \Big]
    = \max_\theta \left[
        h_2 \!\left(\frac{1+|\cos(\theta/2 + \phi)|}{2}\right)
        - h_2 \!\left(\frac{1+|\cos(\theta/2)|}{2}\right)
    \right] = \Delta S_{\text{max}}.
    \label{eq:generic_entropy_bound}
\end{equation}

\textbf{Infinitesimal entangling rate and ancilla independence.} For small angle $\phi$, we can expand $S(\theta + 2 |\phi|) = S(\theta) + 2 |\phi| S'(\theta) + \mathcal{O}(\phi^2)$, and thus the bound is linear in $\phi$~\cite{Dur2001},
\begin{equation}
    \Delta S_{\text{max}} = 2 |\phi| \max_\theta |S'(\theta)| + \mathcal{O} (\phi^2),
\end{equation}
where $\max_\theta |S'(\theta)| \approx 0.4780683222$. Clifford transformations within subsystems reduce any Pauli gadget crossing the bipartition to a two-qubit interaction, but the remaining qubits may act as entangled ancillas. For self-inverse product Hamiltonians, ancillas provably do not increase the maximal infinitesimal entangling rate~\cite{Wang2003}, extending the linear bound to arbitrary multi-qubit states. Numerical evidence suggests that this ancilla independence persists at finite angles~\cite{Leifer2003}, supporting the extension of Eq.~\eqref{eq:generic_entropy_bound} to multi-qubit states.

\section{Entanglement entropy change for Pauli gadgets on a graph state}

Consider a graph state $\ket{G}$ and a set of Pauli gadgets $U_i = e^{- i \varphi_i P_i / 2}$. The purpose is to determine the change in entanglement entropy after the application of all Pauli gadgets. The graph state $| G \rangle$ with chosen subsystems $A$ and $B$ can always be expressed in the following form:
\begin{equation}
  (C_A \otimes C_B) | G \rangle = | \tilde{G} \rangle = | \Phi_D \rangle_{A_e,
  B_e} \otimes | \mathbf{0} \rangle_{A_l} \otimes | \mathbf{0}
  \rangle_{B_l},
\end{equation}
where $C_A, C_B$ are the Clifford unitaries supported on subsystem $A$ or $B$, $| \Phi_D \rangle$ is a maximally entangled state with entanglement entropy $S_0$ and Schmidt rank $D = 2^{S_0}$, regions $A_e$ and $B_e$ are entangled degrees of freedom (subregions of $A$ and $B$ with exactly $D$ qubits each), and regions $A_l$ and $B_l$ (the rest of $A$ and $B$) are only consisting of $| 0 \rangle$ ``ancillas''. Note that $\ket{\tilde G}$ is not necessarily a graph state, but it is equivalent to one, up to single-qubit Cliffords.

The Clifford unitaries $C_A, C_B$ can also be used to transform all Pauli gadgets $U_i = e^{- i \varphi_i P_i / 2}$,
\begin{equation}
  (C_A \otimes C_B) U_i (C_A \otimes C_B)^{\dag} = (C_A \otimes C_B) e^{- i
  \varphi_i P_i / 2} (C_A \otimes C_B)^{\dag} = e^{- i \varphi_i  \tilde{P}_i
  / 2} .
\end{equation}
In the following, we will drop the tilde from the notation, assuming that we always work with the transformed operators and states. Importantly, the unitary transformations performed here are localised to $A$ and $B$, and therefore do not change the entanglement entropy.

\subsection{One gadget}

We begin by applying one gadget to the graph state,
\begin{align}
  \ket{\psi} & = e^{- i \varphi_1 P_1 / 2} \ket{G}
  \\
  & = \left( \cos \frac{\varphi_1}{2} I - i \sin \frac{\varphi_1}{2} P_1   \right) (| \Phi_D \rangle_{A_e, B_e} \otimes | \mathbf{0} \rangle_{A_l} \otimes | \mathbf{0} \rangle_{B_l})
  \\
  & = \cos \frac{\varphi_1}{2}  | G \rangle - i \sin \frac{\varphi_1}{2} 
  (I \otimes P_1^{B_e} (P_1^{A_e})^T) | \Phi_D \rangle_{A_e, B_e} \otimes
  P_1^{A_l} | \mathbf{0} \rangle_{A_l} \otimes P_1^{B_l} | \mathbf{0}
  \rangle_{B_l}. \label{eq:statedecomp}
\end{align}
The last equality follows from the property of a maximally entangled state, $(A \otimes B) | \Phi_D \rangle = (I \otimes B A^T) | \Phi_D \rangle = (A B^T \otimes I) | \Phi_D \rangle$.

$P_1^{A_l}$ applied on $| \mathbf{0} \rangle_{A_l}$ will either leave the
qubits in $| \mathbf{0} \rangle_{A_l}$ (up to a phase factor) if $P_1^{A_l}$ only consists of $I$ and $Z$, or will change it to an orthogonal state $| \mathbf{x} \rangle_{A_l}$ if $P_1^{A_l}$ consists of $X$ or $Y$, where $\mathbf{x} \in \mathbb{F}_2^{| A_l |}$.
Similarly, $P_1^{B_l}$ will either leave $| \mathbf{0} \rangle_{B_l}$ or will change it to some orthogonal state $| \mathbf{y} \rangle_{B_l}$.
Hence, the resulting state is a superposition of the following two terms:
\begin{equation}
    P_1^z \ket{G} = \eta_z (I\otimes R_z) \ket{\Phi_D}_{A_e,B_e} \otimes \ket{\mathbf{x}}_{A_l} \otimes \ket{\mathbf{y}}_{B_l},
\end{equation}
where $z \in \{0,1\}$, $\eta_z$ is a phase factor of $\{1,i,-1,-i\}$, $R_{z=0} = I$ and $R_{z=1}=P_1^{B_e} (P_1^{A_e})^T$. Each term is taken with a corresponding weight, $\cos(\phi_1/2)$ or $-i \sin(\phi_1/2)$.

This form allows us to determine all possible changes in entanglement entropy relatively easily. The entropy remains $S_0$ if for $z=1$ bitstrings $\mathbf{x}$ or $\mathbf{y}$ are $\mathbf{0}$ and $R_z$ is Hermitian. If $R_z$ is not Hermitian for $z=1$, then the entropy decreases to $S_0 -1 + h_2(\frac{1}{2}(1-\sin \varphi_1))$, where $h_2(x) = -x \log x - (1-x) \log (1-x)$ is the binary entropy. Finally, if $\mathbf{x}\neq\mathbf{0}$ and $\mathbf{y}\neq\mathbf{0}$ for $z=1$, then the two branches in the superposition are orthogonal, increasing the entanglement entropy to $S_0 + h_2(\sin^2(\varphi_1/2))$.

Therefore, the change of entanglement entropy for one Pauli gadget applied on a graph state can be bounded from above by $S - S_0 \le h_2(\sin^2(\varphi_1/2))$.

\subsection{Two gadgets}

Consider now two gadgets $U_1$ and $U_2$ applied on the graph state.
The resulting state is a superposition of four terms,
\begin{equation}
    P_2^{z_2} P_1^{z_1} \ket{G} = \eta_{\mathbf{z}} (I\otimes R_{\mathbf{z}}) \ket{\Phi_D}_{A_e,B_e} \otimes \ket{\mathbf{x}}_{A_l} \otimes \ket{\mathbf{y}}_{B_l},
\end{equation}
where the bitstrings are $\mathbf{x} = z_1  \mathbf{x}_1 \oplus z_2  \mathbf{x}_2$, $\mathbf{y} = z_1  \mathbf{y}_1 \oplus z_2  \mathbf{y}_2$, $\mathbf{z}=\{z_1, z_2\}$, and $\oplus$ is the addition modulo 2. $\mathbf{x}_i$ is the bitstring after applying $P_i^{A_l}$ on $\ket{\mathbf{x}_i}_{A_l}$, and similarly for $\mathbf{y}_i$, while $\eta_{\mathbf{z}}$ is a phase factor one obtains after transforming the bitstrings.

Calculation of the entropy is now slightly more involved, but can be done by forming a $4D \times4D$ matrix of amplitudes, with submatrices of size $D\times D$ given by
\begin{equation}
  M_{\mathbf{u}, \mathbf{v}} = \sum_{\mathbf{z} \,:\, \mathbf{x}
  (\mathbf{z}) = \mathbf{u}, \mathbf{y} (\mathbf{z}) = \mathbf{v}}
  \beta_{\mathbf{z}} R_{\mathbf{z}} ,
\end{equation}
where $\beta_{\mathbf{z}} = \eta_{\mathbf{z}}  (- i)^{\sum_i z_i}  \prod_i
\cos^{1 - z_i} (\varphi_i/2) \sin^{z_i} (\varphi_i/2)$ is the amplitude of the term corresponding to
$\mathbf{z}$ in the superposition. Singular values of the matrix $M$ are, after normalization, the Schmidt coefficients of the final state, $\{\sigma_r(M)/\sqrt{D}\}$, where $\sigma_r(\cdot)$ designates the $r$-th singular value of the matrix. Entanglement can thus be calculated as
\begin{equation}
    S=-\sum_{r=1}^{4D} \frac{\sigma_r(M)^2}{D} \log_2 \frac{\sigma_r(M)^2}{D}.
\end{equation}

We find that for two gadgets, entanglement entropy can have 23 distinct forms. This classification arises because two gadgets generate four stabilizer-state branches, whose interference is constrained by the stabilizer structure and their overlaps across the bipartition. Most can be thought of as a straightforward generalization of the one-gadget results, where the entanglement entropy is bounded from above by an additive bound $S \le S_0 + h_2(\sin^2(\varphi_1/2)) + h_2(\sin^2(\varphi_2/2))$. However, we also find scenarios when this bound is sometimes violated: the Pauli gadgets can be fused into one gadget with phase $\varphi_1 \pm \varphi_2$. This happens when $P_1 P_2 \ket{G} = \pm \ket{G}$, i.e.\@ $P_1$ and $P_2$ differ only by a stabilizer of the graph state. The resulting entropy is a descendant from the one-gadget case, i.e. $S_0 -1 + h_2(\frac{1}{2}(1-\sin (\varphi_1 \pm \varphi_2)))$ or $S_0 + h_2(\sin^2((\varphi_1 \pm \varphi_2)/2))$; the latter can violate the bound of $S_0 + h_2(\sin^2(\varphi_1/2)) + h_2(\sin^2(\varphi_2/2))$. For a randomly chosen set of Paulis, gadget fusion is exponentially suppressed.

\subsection{General scheme for multiple gadgets}

We now apply $N$ number of gadgets $U_i$ to the graph state $| G \rangle$. The result is a superposition of $2^N$ terms of the form
\begin{equation}
  \underbrace{P_N^{z_N} \ldots P_2^{z_2} P_1^{z_1}}_{Q_{\mathbf{z}}} | G \rangle = \eta_{\mathbf{z}} (I \otimes R_{\mathbf{z}}) | \Phi_D \rangle_{A_e, B_e} \otimes | \mathbf{x} \rangle_{A_l} \otimes | \mathbf{y} \rangle_{B_l},
  \label{eq:superposition}
\end{equation}
where $\eta_{\mathbf{z}} \in \{ 1, i, - 1, - i \}$ is the phase factor of the bitstring transformation associated with a specific choice of $\mathbf{z} = \{ z_1, z_2, \ldots, z_N \}$.

Similarly to the two-gadget case, we create the matrix $M$ of size $2^N\times 2^N$ with submatrices
\begin{equation}
  M_{\mathbf{u}, \mathbf{v}} = \sum_{\mathbf{z} \,:\, \mathbf{x}
  (\mathbf{z}) = \mathbf{u}, \mathbf{y} (\mathbf{z}) = \mathbf{v}}
  \beta_{\mathbf{z}} R_{\mathbf{z}} ,
\end{equation}
where again $\beta_{\mathbf{z}} = \eta_{\mathbf{z}}  (- i)^{\sum_i z_i}  \prod_i \cos^{1 - z_i} (\varphi_i/2) \sin^{z_i} (\varphi_i/2)$ is the amplitude of the term corresponding to $\mathbf{z}$ in the superposition.
Matrix $M$ is directly related to the reduced density matrix as $\rho_A = M M^\dagger / D$. The entanglement entropy is given by
\begin{equation}
    S=-\sum_{r=1}^{2^N D} \frac{\sigma_r(M)^2}{D} \log_2 \frac{\sigma_r(M)^2}{D}.
\end{equation}

\textbf{Upper bound.} We can, however, simplify this formula considerably by bounding it from above. Consider dephasing $\rho_A = M M^\dagger / D$ into blocks of constant $\mathbf{u}$,
\begin{equation}
    \Delta_{\mathbf{u}}(\rho_A) = \bigoplus_{\mathbf{u}} \frac{1}{D} \sum_{\mathbf{v}} M_{\mathbf{u},\mathbf{v}} M_{\mathbf{u},\mathbf{v}}^\dagger = \sum_{\mathbf{u},\mathbf{v}} p_{\mathbf{u},\mathbf{v}} \rho_{\mathbf{u},\mathbf{v}},
\end{equation}
where $\Delta_{\mathbf{u}}(\rho_A)$ is the reduced density matrix after dephasing, $p_{\mathbf{u}, \mathbf{v}} = \| M_{\mathbf{u}, \mathbf{v}} \|^2_F / D$ can be regarded as a mixture probability ($\sum_{\mathbf{u}, \mathbf{v}} p_{\mathbf{u}, \mathbf{v}}=1$), and $\rho_{\mathbf{u}, \mathbf{v}} = \ket{\mathbf{u}} \bra{\mathbf{u}} \otimes M_{\mathbf{u}, \mathbf{v}} M_{\mathbf{u}, \mathbf{v}}^\dagger / \| M_{\mathbf{u}, \mathbf{v}} \|^2_F$ is a normalized density matrix supported in the $\mathbf{u}$-block. Now, using the fact that dephasing cannot decrease entropy~\cite{Nielsen2010}, and using the mixture entropy bound, $S(\sum_i p_i\rho_i) \leq H(\{p_i\}) + \sum_i p_i S(\rho_i)$~\cite{Nielsen2010}, we find that the entropy of the state is bounded as
\begin{equation}
    S \le S(\Delta_{\mathbf{u}}(\rho_A)) \le H(\{p_{\mathbf{u}, \mathbf{v}}\}) + \sum_{\mathbf{u}, \mathbf{v}} p_{\mathbf{u}, \mathbf{v}} S(\rho_{\mathbf{u}, \mathbf{v}}),
\end{equation}
where $H(\{x_i\}) = -\sum_i x_i \log_2 x_i$ is Shannon entropy. Finally, we use the fact that $S(\rho_{\mathbf{u}, \mathbf{v}})$ can be calculated from the singular values of submatrix $M_{\mathbf{u}, \mathbf{v}}$, as $S(\rho_{\mathbf{u}, \mathbf{v}}) = H(\{ \sigma_r(M_{\mathbf{u}, \mathbf{v}})^2 / \| M_{\mathbf{u}, \mathbf{v}} \|^2_F \})$, where now index $r\in\{1,...,D\}$, and that this expression is bounded from above by $\log_2 D = S_0$. This yields the formula
\begin{equation}
    S \le S_0 + H(\{p_{\mathbf{u}, \mathbf{v}}\}).
\end{equation}

Let us now assume that either none of the $\mathbf{z}$-branches combine in the superposition in Eq.~\eqref{eq:superposition} (i.e., $Q_{\mathbf{z}} \ket{G} \neq \gamma_{\mathbf{z}, \mathbf{z}'} Q_{\mathbf{z}'} \ket{G}$, where $\gamma_{\mathbf{z}, \mathbf{z}'} \in \{1, i, -1, -i\}$), or they do so not coherently (i.e., $Q_{\mathbf{z}} \ket{G} = \gamma_{\mathbf{z}, \mathbf{z}'} Q_{\mathbf{z}'} \ket{G}$, but $\text{Re}[\beta_{\mathbf{z}}^* \beta_{\mathbf{z}'} \gamma_{\mathbf{z}, \mathbf{z}'}] = 0$). In this case, the block probabilities $p_{\mathbf{u}, \mathbf{v}}$ are sums of the independent $\mathbf{z}$-branch distributions $|\beta_{\mathbf{z}}|^2$, so that $H(\{p_{\mathbf{u}, \mathbf{v}}\}) \le H(\{|\beta_{\mathbf{z}}|^2\}) = \sum_i h_2(\sin^2(\varphi_i/2))$. Thus, we have a stricter upper bound on entanglement,
\begin{equation}
    S \le S_0 + \sum_i h_2\!\left(\sin^2 \frac{\varphi_i}{2}\right),
    \label{eq:entropy_bound}
\end{equation}
where the sum is taken over only the gadgets that cross the bipartition. 
Note that in the sum, we can also discard the gadgets that, after transforming using a stabilizer of the state, are supported on one side of the bipartition.

Violation of the bound in Eq.~\eqref{eq:entropy_bound} can only happen if $\mathbf{z}$-branches combine coherently, that is, $Q_{\mathbf{z}} \ket{G} = \gamma_{\mathbf{z}, \mathbf{z}'} Q_{\mathbf{z}'} \ket{G}$ and $\text{Re}[\beta_{\mathbf{z}}^* \beta_{\mathbf{z}'} \gamma_{\mathbf{z}, \mathbf{z}'}] \neq 0$. This means we can find a minimal set $\mathcal{T}$ of gadget indices (ordered according to how gadgets are applied on the state) for which
\begin{align}
    \prod_{i\in\mathcal{T}} P_i &= \pm \text{Stab}(\ket{G}) \quad \text{for even number of gadgets in }\mathcal{T},\\
    \prod_{i\in\mathcal{T}} P_i &= \pm i\, \text{Stab}(\ket{G}) \quad \text{for odd number of gadgets in }\mathcal{T}.
\end{align}
We term this situation ``coherent branch interference''. For example, when $\mathcal{T}$ contains only two gadgets, both in the same commuting region, it corresponds to a gadget fusion scenario in the ZX-calculus. Many violations of the entropy bound in Eq.~\eqref{eq:entropy_bound} can be removed by fusing gadgets that differ by a stabilizer of the state.

\textbf{Lower bound.} Assuming no coherent branch interference, the states $Q_{\mathbf z}\ket{G}$ appearing in Eq.~\eqref{eq:superposition} are distinct and orthogonal, and can be completed to the stabilizer-orbit basis $\{Q\ket{G}\}$, with measurement probabilities $|\beta_{\mathbf z}|^2$ (and zero otherwise). We apply the Maassen-Uffink uncertainty relation~\cite{Maassen1988} between this basis and a complete product basis containing the Schmidt vectors $\ket{a_j}\ket{b_j}$ of the final state, for which the measurement entropy is the entanglement entropy $S$ that we wish to bound. Since the graph state has a flat Schmidt spectrum of rank $D=2^{S_0}$, every $Q\ket{G}=(Q_A\otimes Q_B)\ket{G}$ has the same Schmidt coefficients $1/\sqrt D$ and hence overlaps at most $c=1/\sqrt D$ with any product state. Maassen-Uffink therefore gives $S+H(\{|\beta_{\mathbf z}|^2\}) \geq -2\log_2 c = S_0$. Using $H(\{|\beta_{\mathbf z}|^2\})=\sum_i h_2[\sin^2(\varphi_i/2)]$ yields the lower bound for entanglement entropy,
\begin{equation}
    S \ge S_0 - \sum_i h_2\!\left(\sin^2 \frac{\varphi_i}{2}\right).
    \label{eq:lower_entropy_bound}
\end{equation}

\begin{figure}
    \centering
    \includegraphics[width=0.7\linewidth]{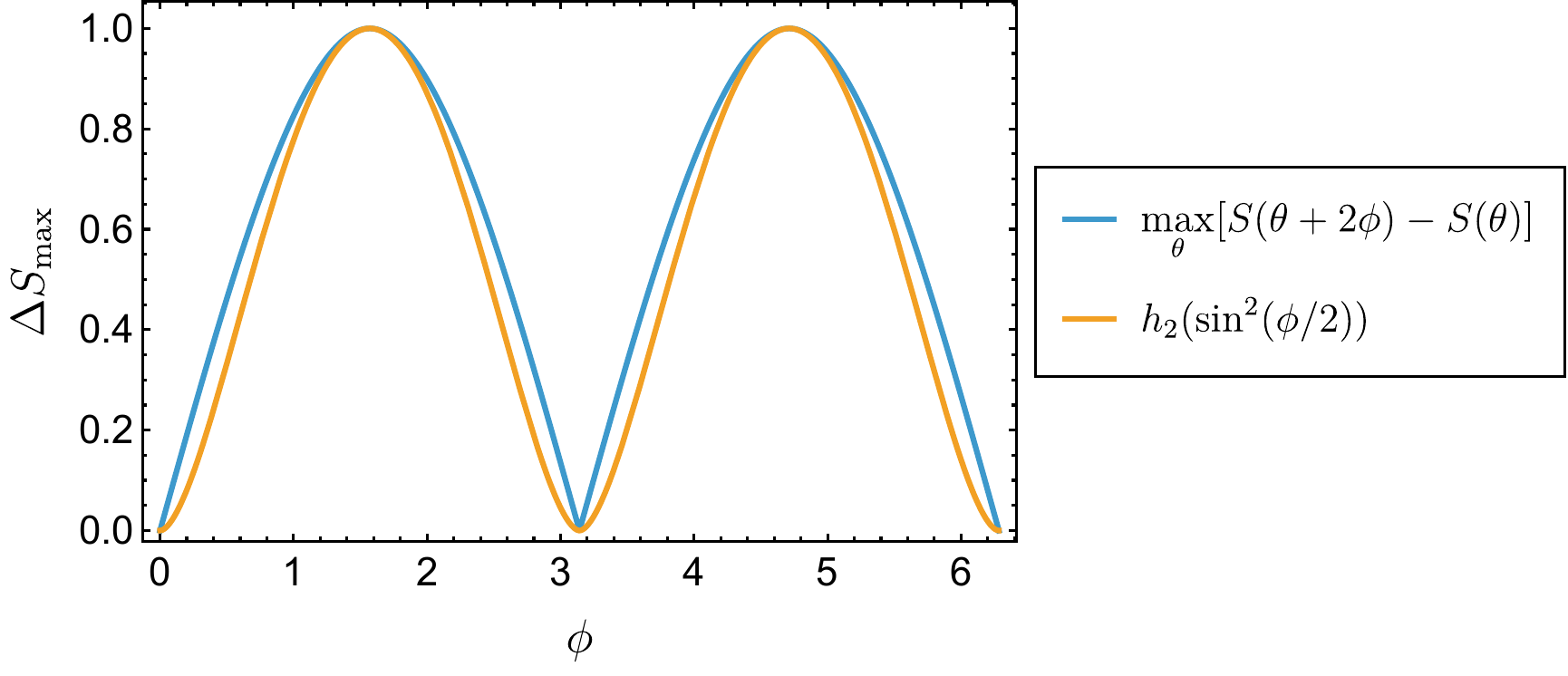}
    \caption{Comparison between a more generic bound from Eq.~\eqref{eq:generic_entropy_bound} and the bound without gadget fusion from Eq.~\eqref{eq:entropy_bound}. Plotted is the maximum change in entanglement entropy, $\Delta S_{\rm max}$, as a function of the Pauli gadget phase $\phi$.}
    \label{fig:generic_bound_comparison}
\end{figure}

\textbf{Comments.} Note that the bounds in Eq.~\eqref{eq:entropy_bound} and \eqref{eq:lower_entropy_bound} are much stricter than the bound in Eq.~\eqref{eq:generic_entropy_bound}, which can be seen in Fig.~\ref{fig:generic_bound_comparison} and by expanding these expressions for small angles $\varphi_i$, which yields an expression proportional to $\varphi_i^2 \log_2 \varphi_i$ in the leading order,
\begin{equation}
    |S - S_0| \le \frac{1}{4} \sum_i \left( \frac{1}{\ln 2} + 2 - 2 \log_2 \varphi_i \right) \varphi_i^2 + \mathcal{O}(\varphi_i^4 \log_2 \varphi_i).
\end{equation}

The von Neumann entropy bounds can be adapted to R\'enyi entropies $S_\alpha$, where $\alpha$ is the R\'enyi index. The upper bound derivation proceeds similarly, with $h_2(x)$ replaced by the R\'enyi binary entropy $h_2^{\alpha}(x) = (1-\alpha)^{-1} \log_2 (x^\alpha + (1-x)^\alpha)$. The lower bound derivation differs slightly, as the R\'enyi Maassen-Uffink relation pairs entropies with conjugate indices, giving $S_\alpha+H_{\alpha'}(\{|\beta_{\mathbf z}|^2\}) \geq -2\log_2 c$, where $1/\alpha+1/\alpha'=2$. As a consequence, the bounds for R\'enyi entropies are not symmetric,
\begin{equation}
    S_0 - \sum_i h_2^{\alpha'}\!\left(\sin^2 \frac{\varphi_i}{2}\right)
    \le
    S_\alpha
    \le
    S_0 + \sum_i h_2^{\alpha}\!\left(\sin^2 \frac{\varphi_i}{2}\right),
    \qquad \frac{1}{\alpha} + \frac{1}{\alpha'} = 2.
\end{equation}

\section{Bounds on magic (log stabilizer extent)}

Let us define the log stabilizer extent~\cite{Bravyi2019, Heimendahl2021, Seddon2021} as
\begin{equation}
    \mathcal{M}(\ket{\psi}) = \log_2 \xi(\ket{\psi}),
\end{equation}
where $\xi(\ket{\psi})$ is a stabilizer extent. $\mathcal{M}$ is a logarithmic version of the magic/non-stabilizerness monotone $\xi$ (analogous to the logarithmic negativity or log robustness). The stabilizer extent is defined as
\begin{equation}
    \xi(\ket{\psi}) = \min_{\ket{\psi}=\sum_i \alpha_i \ket{S_i}} (\sum_i |\alpha_i|)^2,
\end{equation}
where the minimum is taken over all possible stabilizer decompositions of the state, $\ket{S_i}$ are the stabilizer states in the decomposition, and $\alpha_i$ are their coefficients.

We aim to bound $\mathcal M$ additively from above. For this purpose, let us first find the maximal increase in $\mathcal M$ after application of one Pauli gadget on an arbitrary state. Take an optimal decomposition of the original state
\begin{equation}
\ket{\psi} = \sum_j \alpha_j\ket{S_j}, \qquad \sum_j|\alpha_j|=\sqrt{\xi_0}.
\end{equation}
We now examine $e^{i\varphi P/2}\ket{S_j}$ for each stabilizer state.
After absorbing signs, global phases, and possibly multiplying by the Clifford $P$, we may write
\begin{equation}
  e^{i\varphi P/2} \ket{S_j} = M \ket{S_j} + i m P \ket{S_j},
\end{equation}
where $M = \max\{|\cos(\varphi/2)|, |\sin(\varphi/2)|\}, m = \min\{|\cos(\varphi/2)|, |\sin(\varphi/2)|\}$.
We also notice that the state $\ket{T_j} = (\ket{S_j}+i P\ket{S_j})/\sqrt{2}$ is a stabilizer state. Therefore, we can decompose
\begin{equation}
  M \ket{S_j}+i m P \ket{S_j} = (M-m) \ket{S_j} + \sqrt{2} m \ket{T_j}.
\end{equation}
The stabilizer extent, assuming this is the optimal decomposition for each term, is
\begin{equation}
  \xi = \xi_0 \left(M + (\sqrt{2} - 1) m\right)^2,
\end{equation}
which is the upper bound for the change in stabilizer extent in general.

This generalizes readily to the graph-state-and-Pauli-gadgets normal form, where the upper bound for the log stabilizer extent is a sum of contributions from gadgets $\mathcal{O}$ whose supports are not stabilizers of the state,
\begin{equation}
  \mathcal{M}(\ket{\psi})
  \le 2\sum_{j \in \mathcal{O}}
  \log_2 \! \left(  \max(|\sin\frac{\varphi_j}{2}|,|\cos\frac{\varphi_j}{2}|) \right. 
  \left. + (\sqrt2-1) \min(|\sin\frac{\varphi_j}{2}|,|\cos\frac{\varphi_j}{2}|)  \right).
\end{equation}
The expression, after a bit of algebra, simplifies to
\begin{equation}
    \mathcal{M}(\ket{\psi}) \le 2 \sum_{j \in \mathcal{O}}
    \log_2\! \left[ \sqrt{1-|\sin \varphi_j|} + \sqrt{1-|\cos \varphi_j|} \right]\!.
\end{equation}

\section{Quantum circuits used for numerical tests}

\begin{figure}[htb]
    \centering
    \includegraphics[width=\columnwidth]{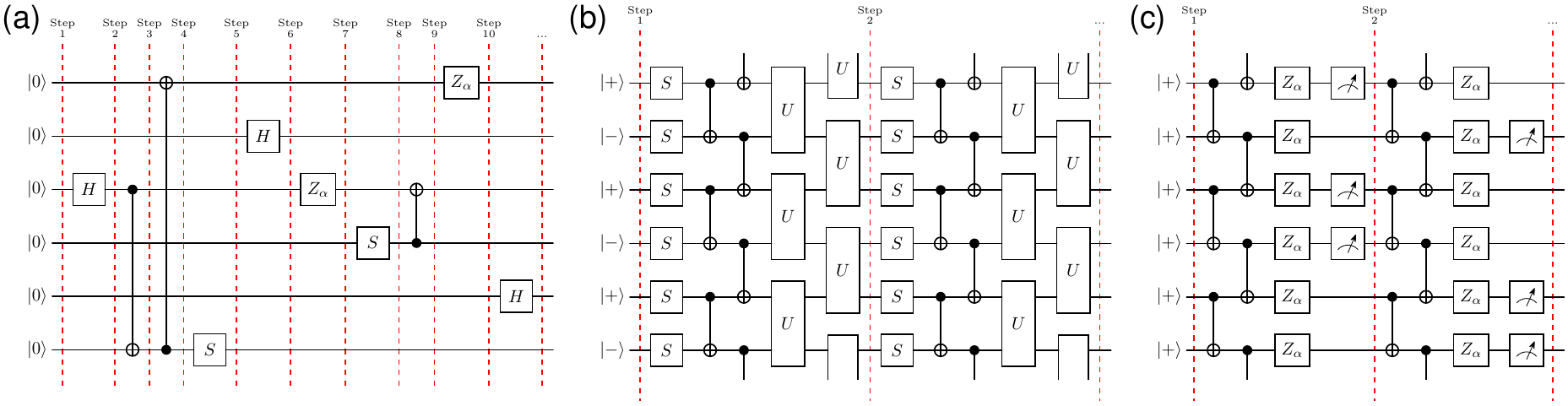}
    \caption{Examples of quantum circuits used for the numerical tests: (a)~random circuit with Clifford gates and small-phase Z-rotations, (b)~circuit combining Trotterized Hamiltonian evolution with Clifford layers, and (c)~random circuit with mid-circuit measurements, where we postselect on $\ket{0}$. $U$ is a Trotterized Hamiltonian evolution $\exp{(-i\,dt\,\mathcal{H})}$, where $\mathcal{H}$ is a generic Hamiltonian.}
    \label{fig:quantum_circuit}
\end{figure}

This appendix provides the explicit circuit constructions used in the numerical tests reported in the main text, for the convenience of the reader. Throughout, we consider a chain of $L$ qubits with periodic boundary conditions. Fig.~\ref{fig:quantum_circuit} presents the three circuits considered here: the random unitary circuit with Clifford gates and small-phase Z-phase gates, circuit combining Trotterized Hamiltonian evolution and Clifford layers, and monitored random circuit.

\end{document}